Atomic Transition Probabilities for Transitions of Si I and Si II and the Silicon Abundances of Several Very Metal-Poor Stars[1]

E. A. Den Hartog[2], J. E. Lawler[2], C. Sneden[3], I. U. Roederer[4,5] & J. J. Cowan[6]

[2]Department of Physics, University of Wisconsin-Madison, 1150 University Ave, Madison, WI 53706; eadenhar@wisc.edu; jelawler@wisc.edu
[3]Department of Astronomy and McDonald Observatory, University of Texas, Austin, TX 78712; chris@verdi.as.utexas.edu
[4]Department of Astronomy, University of Michigan, 1085 S. University Ave., Ann Arbor, MI 48109, iur@umich.edu
[5]Joint Institute for Nuclear Astrophysics – Center for the Evolution of the Elements (JINA-CEE)
[6]Homer L. Dodge Department of Physics and Astronomy, University of Oklahoma, Norman, OK 73019; jjcowan1@ou.edu

ORCIDS:

| | |
|---|---|
| E. A. Den Hartog: | 0000-0001-8582-0910 |
| J. E. Lawler: | 0000-0001-5579-9233 |
| C. Sneden: | 0000-0002-3456-5929 |
| I. U. Roederer | 0000-0001-5107-8930 |
| J. J. Cowan | 0000-0002-6779-3813 |

Abstract

We report new measurements of branching fractions for 20 UV and blue lines in the spectrum of neutral silicon (Si I) originating in the $3s^23p4s$ $^3P^o_{1,2}$, $^1P^o_1$ and $3s3p^3$ $^1D^o_{1,2}$ upper levels. Transitions studied include both strong, nearly pure LS multiplets as well as very weak spin-forbidden transitions connected to these upper levels. We also report a new branching fraction measurement of the $^4P_{1/2} - {^2P^o_{1/2,3/2}}$ intercombination lines in the spectrum of singly-ionized silicon (Si II). The weak spin-forbidden lines of Si I and Si II provide a stringent test on recent theoretical calculations, to which we make comparison. The branching fractions from this study are combined with previously reported radiative lifetimes to yield transition probabilities and log($gf$)s for these lines. We apply these new measurements to abundance determinations in five metal-poor stars.

1. Introduction

Silicon is one of the most abundant elements in the solar system and plays an important role in many astrophysical environments. With its high abundance and relatively low ionization potential it is a significant source of electrons in the interior of cool stars and contributes significantly to the interior opacity in solar-type stars (Amarsi & Asplund 2017). Because silicon is abundant and nonvolatile, it is often used as a reference element to reconcile the absolute scales of meteoritic (e.g. Lodders, Palme & Gail 2009) and solar photospheric abundances (e.g. Asplund et al. 2009). Emission line ratios of Si II, and in particular the ratio of weak resonance lines ($3s^23p\ ^2P^o - 3s3p^2\ ^2D$) and weak intercombination lines ($3s^23p\ ^2P^o - 3s3p^2\ ^4P$), are potentially useful as a plasma diagnostic because of their sensitivity to temperature and density (e.g. Bautista et al. 2009). Silicon-burning, in which $^{28}Si$ is converted to $^{56}Ni$ in a series of successive alpha captures, is the final phase of fusion reactions in the interior of massive stars. Fusion reactions involving elements heavier than $^{56}Ni$ are endothermic and thus not spontaneous. After a brief period (approximately one earth day) of Silicon-burning, the core of a massive star collapses and may explode to release more energy as a Type II supernova.

Motivation for the current study lies in the desire to better understand stellar nucleosynthesis. Records of the "means of production" by which the elements came into being in the earliest epoch of our Galaxy are written into the abundance patterns of the oldest, metal-poor stars in the halo of the Milky Way. Here can be found evidence of the early births, short lives and violent deaths of the first massive stars. Before these abundance patterns can be decoded to gain deeper understanding of the history of nucleosynthesis, we must first be able to determine the abundances of the elements with accuracy and precision. This requires both accurate atomic data and realistic stellar models. As an α–capture element[7], trends of abundance ratios such as [Si/Fe][8] with metallicity yield insight into stellar nucleosynthesis and the chemical evolution of the Galaxy. In an earlier study of the heaviest α–element, Ca, we made detailed comparison between new and published experimental transition probabilities for Ca I and modern theory (Den Hartog et al. 2021). In the present study we make similar comparison with improved transition probabilities for lines of Si I and Si II.

In §2 below, we present a discussion of our measurement method including a description of a new radiometric calibration technique for our high-resolution spectrometer. We present our transition probabilities for 20 lines of Si I and two Si II intercombination lines in §3 along with comparison to the best experimental and theoretical results from the literature. In §4 we apply the new data to derive Si abundances in five warm, very metal-poor main-sequence stars.

---

[7] Formally an α-element is one whose dominant isotope is composed of multiple $^4He$ nuclei. The major natural isotopes of Si (Z = 14) are $^{28}Si$ (92.191% in the solar system), $^{29}Si$ (4.645%), $^{30}Si$ (3.037%) (Meija et al. 2016). For astrophysical purposes, Si is pure $^{28}Si$. Since the minor isotopes of Si collectively contribute only 7.6% to the Si elemental abundance, they will not contribute significantly in solar and stellar optical spectra.

[8] We use standard abundance notations. For elements X and Y, the relative abundances are written [X/Y] = $\log_{10}(N_X/N_Y)_{star} - \log_{10}(N_X/N_Y)_\odot$. For element X, the "absolute" abundance is written $\log_{10} \varepsilon(X) = \log_{10}(N_X/N_H) + 12$. Metallicity is defined as the stellar [Fe/H] value. We adopt the Solar reference abundances from Asplund (2009).

## 2. Emission Branching Fractions

The technique of combining radiative lifetimes from laser-induced fluorescence measurements with emission branching fractions (BFs) measured using high-resolution spectrometers is now the standard method for measuring transition probabilities, or Einstein *A*-values, with efficiency and accuracy (e.g. Lawler et al. 2009). The BF for a transition between an upper level *u* and a lower level *l* is given by the ratio of its *A*-value to the sum of the *A*-values for all transitions associated with *u*, which is the inverse of the radiative lifetime, $\tau_u$. Thus the radiative lifetime, $\tau_u$, provides the absolute scale when converting a BF to an *A*-value. For the purposes of measuring BFs, it can also be expressed as the ratio of relative emission intensities *I* (in any units proportional to photons/time) for these transitions:

$$BF_{ul} = \frac{A_{ul}}{\sum_l A_{ul}} = A_{ul}\tau_u = \frac{I_{ul}}{\sum_l I_{ul}}. \tag{1}$$

BFs, by definition, sum to unity. In order to assure the correct normalization, it is therefore important when measuring BFs to account for all possible decay paths from an upper level. If some weak transitions cannot be measured, these "residual" BFs need to be estimated from theory and accounted for in the total decay rate. If the sum is over significantly less than the full complement of lines, then one has a branching ratio (BR).

In order to avoid line blends, a high-resolution spectrometer is usually required to measure the emission branching fractions unless the spectrum is very sparse. Often a Fourier transform spectrometer (FTS) is used as these instruments have many advantages, including high-resolution, broad spectral coverage and excellent absolute wavenumber accuracy. FTS instruments have one significant disadvantage in that the quantum noise in the spectrum gets spread evenly throughout the spectrum. This "multiplex" noise results in weak lines being swamped in the noise from the strong lines in the spectrum. To overcome the multiplex noise the lamp current is often increased to the point that strong lines in the spectrum are affected by optical depth, which in turn results in inaccurate BFs. Corrections for optical depth can be made, but if the corrections are large they lead to increased uncertainties.

In the current study, BFs in Si I and II have been determined from spectra recorded with the University of Wisconsin (UW) high-resolution echelle spectrograph. This instrument is described in detail in Wood & Lawler (2012). As a dispersive instrument, it does not have multiplex noise and is much better-suited than an FTS for measurement of weak lines while keeping source currents low and avoiding significant self-absorption on the strong transitions. It is a 3-m cross-dispersed echelle spectrograph with broad spectral coverage, resolving power $R \geq 250{,}000$ and a 4 Mpixel CCD detector. The spectra are two-dimensional CCD images containing multiple grating orders, with the high-resolution of each grating order running in one direction and the orders arranged side-by-side in the other dimension. The cross-disperser utilizes a prism to separate the orders, so the orders are further apart at lower wavelength and get increasingly closer together at higher wavelengths. In the far-ultraviolet (far-UV) one CCD frame covers approximately 150 nm in the low resolution direction and three overlapping frames are required to capture an entire grating order in the high-resolution direction. The usual mode of operation would be to acquire five overlapping frames for each UV spectrum, to provide some redundancy and check for source drifts. However, the wavelengths of transitions from the upper levels in the current study are such that all transitions from each level can be studied with a single grating setting. This serendipitous coincidence of line placement means that there is no need to combine frames with different grating settings, eliminating the contribution to the uncertainty that such combining generates.

The optical sources used for generating the Si I, II spectra are commercially manufactured Si-Ne and Si-Ar hollow cathode lamps (HCLs). Each CCD frame recorded is accompanied by a continuum lamp spectrum recorded after the frame, from which a relative radiometric calibration for that frame is determined. In the current study a deuterium ($D_2$) lamp is used as the calibration light source. The only change made between these two recordings is the angle of a steering mirror on a kinematic mount. Beyond this mirror light from each lamp encounters the same optical path. Table 1 lists all spectra recorded for the current study of Si II and Si I BFs. The spectra are analyzed by taking a numerical integral of each line across the width of the grating order in which it is found and dividing that by an integral of the $D_2$ lamp intensity at the same CCD position. The relative irradiance of the $D_2$ lamp can be used to put all lines on the same relative scale. These radiometrically calibrated intensities are then converted to BFs using Equation 1.

Multiple spectra are taken of our primary source, the Si-Ne HCL, over a range of currents between 3 mA and 32 mA. A range of lamp currents is used to check for evidence of self-absorption on the strongest lines of Si I. Self-absorption becomes apparent by studying the BR of a weaker line from the same upper level compared to a strong line that connects to the lowest term. If self-absorption is present on the strong transition this BR will increase with increasing lamp current. We see some evidence of minor self-absorption on three strong Si I lines that connect to the ground term. These have small corrections applied based on the extrapolation of the BR to zero current. The largest of these extrapolations is only 2% lower than the BR measured on the lowest current spectrum.

## 2.1 Detector-based Radiometric Calibration

A continuum lamp is required for the calibration of the echelle spectrometer in order to capture the rapidly changing instrument sensitivity along the grating orders due to the $sinc^2$ blaze envelope of the grating. However, the calibration in the low resolution direction, which changes slowly as a function of wavelength, can be achieved by some other means and then transferred onto the $D_2$ source. For this project we have chosen to use a National Institute of Standards and Technology (NIST) calibrated photodiode detector as our standard. Switching to a detector-based standard from a source-based standard has the advantage that the detector will remain stable for many years, whereas lamp sources age both with shelf life and with usage. UV damage to the window causes changes to the radiant output, particularly in the far-UV. The irradiance of the lamp has to be periodically checked against another little-used lamp and then corrections applied, or the lamp must be sent out to be recalibrated at considerable expense. Another motivation for switching to the detector-based calibration is that $D_2$ lamps are only calibrated between 200 nm and 400 nm and the current project required a calibration out to 410 nm. Even the calibrated irradiance above 370 nm requires careful correction in order to use the lamp at high resolution. This is because above 370 nm there are increasing numbers of lines in the $D_2$ lamp spectrum in addition to the continuum radiation. The original irradiance calibration of our lamp was made with a 4 nm bandpass,[9] effectively smoothing over the increasing forest of lines. At high resolution these lines are resolved and care must be taken to use only continuum radiation when calibrating the metal line intensities. For past studies we have estimated corrections such that the corrected irradiance gave the irradiance of the continuum only rather than an average of continuum plus lines, but such corrections introduce additional uncertainty in the calibration.

The detector used in this calibration is a Hamamatsu S2281 silicon photodiode calibrated at NIST over the wavelength range 200 – 1100 nm. The accuracy of this calibration is 1.2 - 0.34 % over the 200 – 410 nm range of the present study. A line source is also required and we have chosen a Hg pen lamp because it has a spectrum sparse enough that only one to a few lines are transmitted through each of the narrowband optical filters employed, as described below. It is also necessary that the source has short term stability over the period of several hours which is the case for the Hg pen lamp. It does not need to have long term stability. Also required for this calibration are several narrowband optical filters which allow a subset of Hg lines through each filter. We have used filters centered at wavelengths of 250 nm, 296 nm, 313 nm, 365 nm, 405 nm and 436 nm. In addition we have used a sharp-cut colored glass filter (Corning 0-56) to block the strong 254 nm light from leaking through the 296 and 313 nm filters. The narrowband filters are ½ inch diameter, and are mounted in a ten position filter wheel for ease and reproducibility of switching from one to the next. One position in the filter wheel is left open with no filter installed to allow unfiltered light from the $D_2$ lamp through.

Figure 1 shows a schematic of the measurement layout. Two lamps are employed, the Hg pen lamp and the $D_2$ lamp, each mounted at one of the positions viewed by the steering mirror on a kinematic mount. Light from either lamp is imaged on the entrance pinhole of the 3-m echelle spectrometer with a focusing mirror. The Hg pen lamp is rotated in its holder such that the pair of capillaries are viewed side-on rather than front on, to limit structure in the image. Light from the source passes through an iris, which limits the cross section of the beam, and then through the filter wheel before reaching the pinhole. When the filter wheel is set to either the 296 nm or 313 nm filter, a two inch square colored glass filter (Corning 0-56 sharp-cut filter) is mounted just in front of the iris (not shown in Figure 1). The calibrated photodiode is moved into the path between the filter wheel and the entrance pinhole to measure the power of light transmitted by each filter. This is done at both the beginning of measurement and then again at the end, to make sure the lamp has remained stable. The photodiode is removed for echelle measurements. A full UV spectrum (three frames) is recorded for light passing through each filter. An unfiltered $D_2$ spectrum is recorded on each frame. Calibrated line intensities are determined for all lines getting through each filter by dividing integrated line intensities by the $D_2$ continuum intensity, using the same analysis software and method as for the Si I,II BFs, as described above. We use the unfiltered $D_2$ spectrum to determine the filtered line intensities so that the $D_2$ intensity removes the sinc$^2$ dependence of the grating order envelope from the intensities but does not remove the effect of the filter bandpass. The calibration of the photodiode is transferred onto the $D_2$ lamp relative irradiance by insisting that the sum of line intensities through each filter be proportional to the photodiode measurement for each filter (in Amps) divided by the responsivity of the photodiode (in Amp/Watt) and divided by the wavenumber of the transition(s) to convert Watts into something proportional to photons/s. The level of reproducibility for this calibration can be seen in Figure 2 which shows two such measurements of the relative $D_2$ irradiance made approximately one month apart. Since the new calibration only extends down to 250 nm, we use a calibration from our windowless Ar mini-arc lamp ($\lambda < 232$ nm) and our little-used D$_2$ lamp transferred to our everyday D$_2$ lamp to bridge the gap between these two calibrations.

---

[9] private communication from Optronics Laboratories

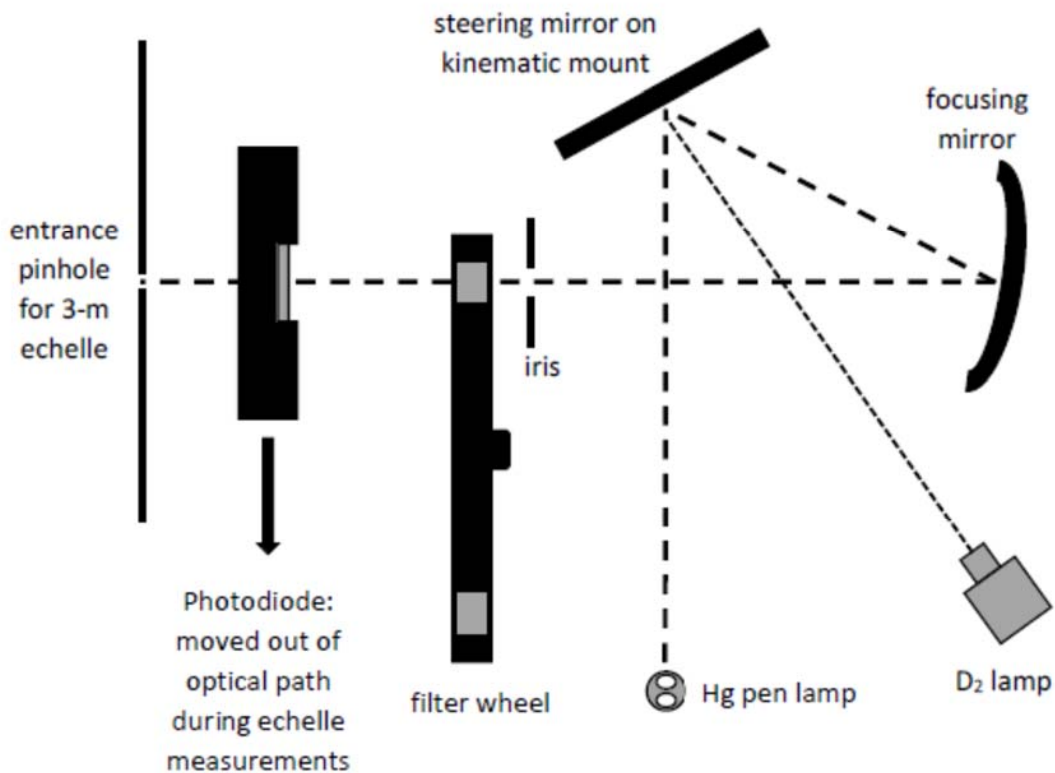

Figure 1. Schematic of the set up for the Hg pen lamp + NIST calibrated photodiode calibration technique.

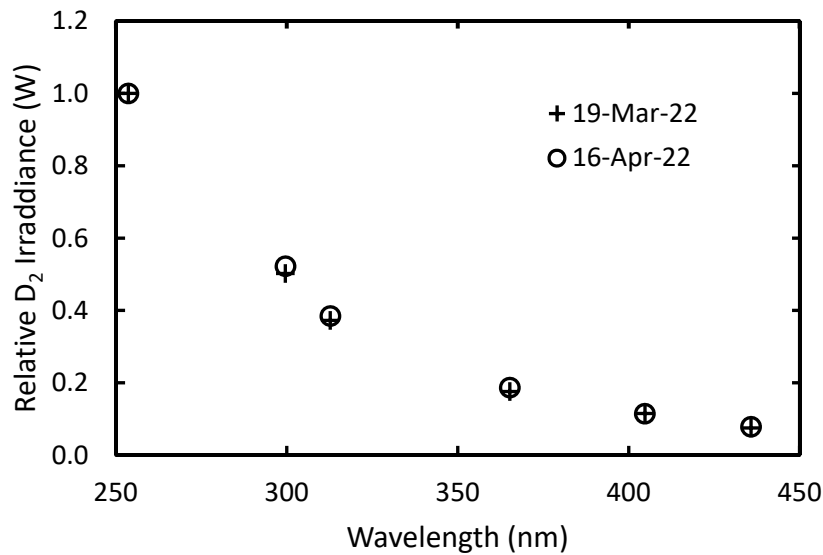

Figure 2. Relative D₂ lamp irradiance between 250 nm and 436 nm as measured on two separate dates using the Hg pen lamp + NIST calibrated photodiode calibration method as described in the text.

It should be mentioned that the Hg pen lamp is not a pure line source but also has a weak continuum component. The paper by Reader, Sansonetti & Bridges (1996) drew our attention to this problem. The weak continuum peaks around 405 nm, but there is also significant continuum associated with the self-absorption on the strong 254 nm line. This continuum contributes to the power measured with the photodiode, but is not accounted for in the filtered line intensity measurements. The problem can be mitigated to some extent by choosing a narrower bandpass for the filter. In the current study we have employed mostly 10 nm bandpass filters, but used a 5 nm bandpass filter at 405 nm where the continuum was strongest. The narrower bandpass reduces the contribution of the continuum relative to the lines. The residual continuum was accounted for by making a measurement of the ratio of line intensity to line+continuum intensity for each filtered spectrum that had some continuum contribution (these were the 250 nm, 365 nm, 405 nm and 436 nm filters). This ratio was then applied as a correction to the photodiode readings in the measurements described above.

We estimate the uncertainty of the calibration to be ~3 – 5% at each point of the curve shown in Figure 2. However, because the D$_2$ irradiance changes smoothly and gradually with wavelength, the uncertainty of the relative calibration between two points on the curve will be less than this estimate and depends on the spacing of the lines being calibrated. A BR for two closely spaced lines, such as the Si II doublet discussed below, will have little contribution to the uncertainty from the calibration whereas lines that are widely separated in wavelength will have a higher contribution to the BR uncertainty. We include a systematic uncertainty of 0.001% per cm$^{-1}$ of wavenumber difference between the line and the dominant line from the upper level as a conservative estimate of uncertainty in the radiometric calibration. This is then added in quadrature to the statistical uncertainty. We estimate the statistical uncertainty as the larger of twice the standard deviation of the weighted mean branching ratio and the inverse of the weighted average signal to noise ratio. The uncertainties of the BRs are then combined using an appropriate error propagation formula to determine the final BF uncertainties.

## 3. Results and Discussion

### 3.1   Si I results

The experimental work on Si I transition probabilities to date has been limited. Garz et al. (1973) determined relative *f*-values for 51 lines between 250 and 800 nm from emission measurements on a wall stabilized arc. They tied these to an absolute scale using radiative lifetimes of Marek (1972). These were later renormalized with new radiative lifetime measurements by Becker et al (1980). Smith et al. (1987; hereafter Sm87) reported experimental BFs or BRs and log(*gf*)s (the log of the level degeneracy multiplied by the oscillator strength) for 108 lines of Si I between 163 and 410 nm. They used a combination of techniques including emission and absorption (Hook) measurements that they tied together using the bowtie method to produce a set of self-consistent relative *f*-values. They chose the beam-foil lifetime measurements of Bashkin et al. (1980) to establish their absolute scale. O'Brian & Lawler (1991, hereafter OL91) measured radiative lifetimes to 5% accuracy for 47 odd-parity levels of Si I and then combined their lifetimes with the BFs of Sm87 for 36 lines originating in 13 of the lower-lying levels that Sm87 studied. Levels above the $3s^23p3d$ $^1P^o_1$ level at 53387 cm$^{-1}$ were deemed by OL91 to have strong infrared branches, and the BFs of Sm87, having only estimated the strength of these transitions, were thought to be less reliable.

There have been a number of theoretical investigations of Si I. Recent studies include the work of Froese Fischer (2005) who used the Breit-Pauli approximation for all levels in Si I up to $3s^23p3d$ $^3D^o$. Savukov (2016; hereafter Sav16) used the configuration-interaction plus many-body-perturbation-theory (CI+MBPT) method to determine transition probabilities, log(*gf*)s and lifetimes for levels of Si I up to the $3s^23p5s$ $^1P^o_1$ level. Wu et al. (2016) used the multi-configuration Dirac-Hartree-Fock (MCDHF) and active space approach to determine levels, hyperfine structure and transition probabilities in Si I up through the $3s^23p4d$ $^3D^o$ levels. Finally, the thesis work of Pehlivan Rhodin (2018; hereafter PR18) used MCDHF method using the GRASP2K package to determine transition probabilities in Si I up through the $3s^23p7s$ and in Si II up through the $3s^27f$ configuration.

Our measured BFs of Si I are presented in Table 2 organized by upper level.[10] Also in this table we compare to a subset of the experimental BFs of Sm87. Note that for several of the weak, spin-forbidden transitions Sm87 only report an upper bound (although what is meant by <0.000 for the $^3D°_{1,2}$ – $^1D_2$ BFs is unclear). In this study, we report the first measurements of these very weak BFs. For lines in common between the two studies, we see an average fractional difference (in the sense (Sm87 – UW)/UW) of +6.0% with a standard deviation of 10.3%. For lines with BFs > 0.01 the average fractional difference is +1.7% with standard deviation of 5.6%.

As a point of reference, we also compare to BFs calculated from LS coupling (also known as Russell-Saunders coupling) theory for the triplet multiplets in Table 2. The upper $3p4s$ $^3P°_1$ and $^3P°_2$ levels at 39760 and 39955 cm$^{-1}$ are nearly pure, with NIST ASD giving the leading percentages as 98 and 99%, respectively. The J=1 level has ~1% mixing with the $3p4s$ $^1P°_1$ level resulting in weak decays to $^1D_2$ and $^1S_0$ lower levels. The upper $3s3p^3$ $^3D°_1$ and $^3D°_2$ levels at 45276 and 45294 cm$^{-1}$ are listed in the NIST ASD as 56% from that configuration and 39% $3pnd$ $^3D°$, but probably have some mixing with nearby $^1P°_1$ and $^1D°_2$ levels, respectively, since both have weak decay to the $3s^23p^2$ $^1D°_2$ level at 6299 cm$^{-1}$. The LS BFs are calculated from relative line strengths tabulated in Appendix I of Cowan (1981). Frequency-cubed scaling is included, and the LS BFs are renormalized to the total multiplet strength as measured in the current study.

Our measured BFs are converted to $A$-values and log($gf$)s following the relations in Thorne et al. (1988),

$$A_{ul} = \frac{BF_{ul}}{\tau_u} \; ; \; \log(gf) = \log\left(\frac{1.499 g_u A_{ul}}{\sigma^2}\right) \quad , \quad (2)$$

where $A_{ul}$ is the transition probability in s$^{-1}$, $\tau_u$ is the radiative lifetime of the upper level in s, $g_u$ is the degeneracy of the upper level, and $\sigma$ is the transition wavenumber in cm$^{-1}$. We use the radiative lifetimes measured previously in our group by OL91 to establish the absolute scale for our BFs. The uncertainty of the $A$-value is the uncertainty of the BF and that of the lifetime added in quadrature. We present $A$-values and log($gf$)s in Table 3. Also in Table 3 we compare to two of the recent theoretical calculations, those by Sav16 and PR18.[11]

Sav16 determined transition probabilities, log($gf$)s and lifetimes only for the low-lying levels of Si I up to the $3s^23p5s$ $^1P°_1$ levels at ~54870 cm$^{-1}$. As such, that study is limited in scope, but achieves relatively high precision on the transitions that it covers by fine-tuning the cavity size, which in turn reduced the basis needed for the lowest states. Sav16 makes detailed comparison to earlier theory of Froese-Fischer (2005) and the experimental *A*-values and radiative lifetimes reported in OL91. We find that we are in good agreement with Sav16 for the 20 transitions studied here even for the weakest transitions down to log(*gf*) < -4. The average fractional difference between our *A*-values (in the sense (Sav16 – UW)/UW) is +1.7% with a standard deviation of 9.7%.

---

[10] Throughout this paper and accompanying tables, Ritz wavelengths and energy levels are taken from the National Institute of Standards and Technology Atomic Spectra Database (NIST ASD; Kramida, Ralchenko & Reader 2021).

[11] We do not make comparison to the best experimental measurements in Table 3. NIST ASD references the results of OL91 (for all but the weakest lines) which combine new lifetime measurements with BFs from Sm87. Our results are not independent from OL91 as we use their lifetimes. We would like to alert the reader that there appears to be an error in the $A$-values and log($gf$)s in the NIST ASD for two of the transitions included in this study: 2443.365 Å and 2452.118 Å. NIST ASD log($gf$)s are +0.32 and -0.53 dex different, respectively, from those found in OL91. This discrepancy is also found in the critical compilation on Silicon by Kelleher & Podobedova (2008).

We also compare in Table 3 to the MCDHF calculations of PR18 who determined transition probabilities for Si I belonging to the even $3s^23p^2$, $3s^23pnp$ (n ≤ 7), and $3s^23pnf$ (n ≤ 6) configurations and to the odd $3s3p^3$, $3s^23pns$ (n ≤ 8), and $3s^23pnd$ (n ≤ 6) configurations. Here we find that the agreement with our measured transition probabilities is also very good, with average fractional difference (in the sense (PR18 – UW)/UW) of -8.5% with a standard deviation of 13.5%. This improves to an average of -3.9% and standard deviation of 10.4% for lines with log(*gf*)>-3. Unlike Sav16, the PR18 study is a comprehensive calculation involving over 100 levels up to 61936 cm$^{-1}$ and more than 1300 transitions ranging in wavelength from 6333 nm in the infrared to 161 nm in the vacuum-UV. As such, it will prove a very valuable resource for astronomers.

The comparisons made in Table 2 and Table 3 are visualized in Figure 3, where we present logarithmic differences (in the sense log(other) – log(UW) versus log(UW)) of the experimental BFs of Sm87 in panel (a) and the log(*gf*)s of PR18 and Sav16 in panels (b) and (c), respectively. In panels (a) and (b) the error bars represent the combined uncertainties added in quadrature. (The uncertainties reported in PR18 are the relative difference between the length and velocity gauges.) Sav16 did not give uncertainties for their *A*-values so no error bars are plotted in panel (c). In panel (a) the point with an arrow beside it is the upper bound quoted in Sm87 for the transition at 4102 Å. The weakest, spin-forbidden transitions in these comparisons are very difficult to measure and to calculate. The level of agreement with recent theory, both with the limited-in-scope but high precision calculations of Sav16, and with the comprehensive calculations of PR18, is very satisfactory.

## 3.2 Si II results

We have remeasured the BF of the very weak spin-forbidden $^4P_{1/2}$ - $^2P^o_{1/2,3/2}$ doublet of Si II at 2334.407 Å and 2350.172 Å, respectively, using the first eight spectra listed in Table 1. Optical depth is not a concern in this measurement because of the weakness of the transitions. This BF had previously been measured in our group and reported in Calamai, Smith & Bergeson (1993, hereafter CSB93). That paper had reported the measurement of the radiative lifetimes of the $^4P_{1/2,3/2,5/2}$ levels as well as the BFs of the $^4P_{1/2}$ level. We use the radiative lifetime of CSB93 to convert our BFs to *A*-values. These are reported in Table 4 along with comparison to the CSB93 measurement. CSB93 report that these lines had signal-to-noise ratios of 10-15 in their spectra whereas we have signal-to-noise ratios ranging from 45 to 200. The radiometric calibration does not significantly contribute to the uncertainty of our BF because of the small wavelength span between the doublet, resulting in an uncertainty that is primarily statistical. The superior signal-to-noise in our spectra explains why our uncertainties are lower than those of CSB93. We also compare to recent theoretical results of PR18 and Wu et al. 2020 in Table 4.

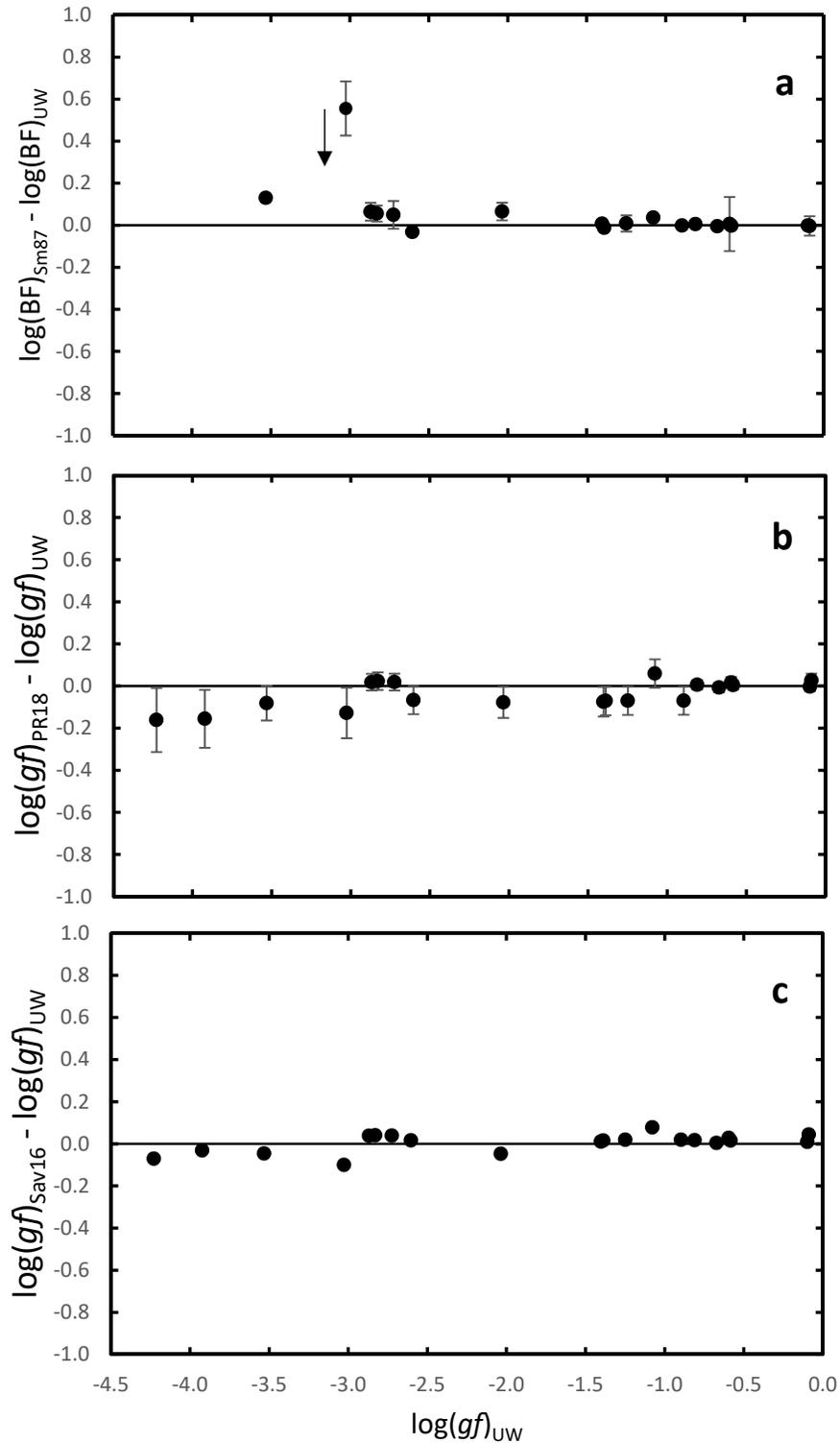

Figure 3. Comparison of log(BF)s or log(*gf*)s of Si I in the present work to those of a) the experimental results of Sm87, b) theoretical MCDHF calculations of PR18, and c) theoretical CI+MBPT calculations of Sav16 versus log(*gf*) from this study. In each figure the horizontal line at 0.0 represents perfect agreement. Error bars represent combined uncertainties where available. See text for further discussion.

CSB93 appear to have used the theoretical BF of Nussbaumer (1977) to convert their lifetime for the $^4P_{3/2}$ level to *A*-values for the $^4P_{3/2}$ - $^2P°_{1/2,3/2}$ doublet at 2328.517 Å and 2344.202 Å. This is not stated clearly in their paper, and in fact they state "Thirty-four measurements of the $^4P_{3/2}$ branching fraction were made. The total uncertainty (systematic and statistical) was about 10% at the 90% level of confidence." This appears to be a typo, and refers to the measurement and uncertainty of the $^4P_{1/2}$ BF. It is stated clearly elsewhere in the paper that a BF was measured for only one level, the $^4P_{1/2}$ level, and the 10% uncertainty mentioned in the quote is not consistent with the 50% uncertainty on the weak branch of the $^4P_{3/2}$ level. We attempted a BF measurement of the $^4P_{3/2}$ - $^2P°_{1/2,3/2}$ doublet at 2328.517 Å and 2344.202 Å, but were unsuccessful. The weaker 2328 Å line of this pair is estimated by the theory of Nussbaumer (1977) and that of Dufton et al. (1991) to be a ~1% branch. Although we saw a weak feature at this wavelength in our higher current Si-Ne spectra, we decided that this feature was a blend with a very weak neon line. There is no observed transition listed at this wavelength in the NIST ASD neon spectrum, but there is a possible Ne II electric dipole transition nearby that obeys parity and J selection rules. Our analysis software looks for these possibilities based on known energy levels of both the metal and buffer gas first and second spectra. We investigated this further by looking at this wavelength in high current Hf-Ne and Hf-Ar spectra taken for a different study. In these spectra we also saw a very weak feature in the Hf-Ne spectra but not in the Hf-Ar spectra, suggesting a neon blend. Unfortunately, switching to a Si-Ar lamp does not help in this case because the other line in the doublet pair, 2344.202 Å, has a known argon blend. We attempted to procure a third commercial HCL with krypton buffer gas which has no potential blends on either line, but were unsuccessful. The most we can say regarding the weak line at 2328.517 Å is that it is less than a 4.5% branch with an upper bound of log(*gf*) < -6.7.

The $^4P$ - $^2P°$ intercombination lines have been part of numerous theoretical investigations of Si II. These lines are allowed E1 transitions due to the mixing of the $3s3p^2$ $^4P$ levels with doublets from the same configuration. The accuracy of calculated radiative rates depend on the accuracy to which the mixing coefficients and the multiplet energy splittings are calculated. Nussbaumer (1977) used the SUPERSTRUCTURE code to calculate radiative parameters from sophisticated configuration interaction wavefunctions. Dufton et al. (1991) significantly improved on those results by including a more extensive set of configurations. These lines were included in the calculations of Froese Fischer (2006) and Tayal (2007) using the MCHF method. Bautista (2009) calculated radiative rates between many configurations using several different approximations and generated a list of recommended log(*gf*)s for transitions among the 15 lowest levels in Si II. Aggarwal & Keenan (2014) used the General-purpose Relativistic Atomic

Structure Package (GRASP()) and the Flexible Atomic Code (FAC) to calculate a large number of radiative parameters and collision strengths in Si II, but estimate ~20% uncertainty on the strong transitions with weak transitions such as these intercombination lines being much more uncertain. PR18 calculate *A*-values for these intercombination lines using the MCDHF method and GRASP2K package with uncertainties based on the relative difference between the length and velocity gauges of ~19% and 12% for the 2334.407 Å and 2350.172 Å lines, respectively. Finally Wu et al. (2020) also used the MCDHF method and the GRASP2K package in their study of Si II. In Figure 4 we make comparison to the experimental results for the BR ($^4P_{1/2}$ - $^2P^o_{3/2}$)/($^4P_{1/2}$ - $^2P^o_{1/2}$) of CSB93 and to the above-mentioned theoretical studies, with the exception of the Aggarwal & Keenan (2014) study. The BR from that study lies significantly off-scale on Figure 4 at 1.37. In this figure the horizontal line is simply a guide to the eye, and lies at the experimental value determined in this study. It can be seen from this figure that the general level of agreement between experiment and theory has improved dramatically over recent decades, undoubtedly owing, at least in part, to rapid increase in computing power. We see particularly excellent agreement between our study and the recent theoretical results of Wu et al. (2020) and PR18 as well as that of Froese Fischer et al. (2006).

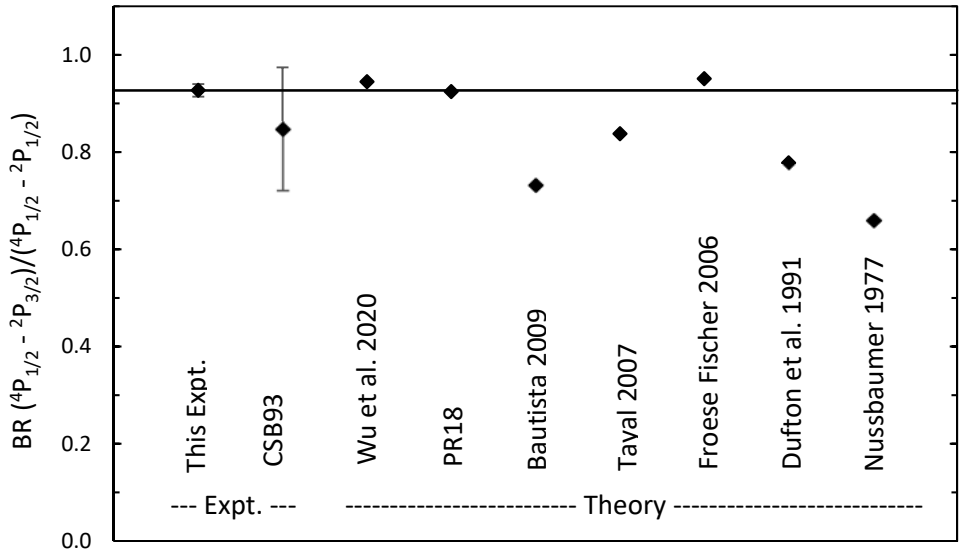

Figure 4. Experimental and theoretical values determined for the BR of the ($^4P_{1/2}$ - $^2P^o_{3/2}$)/($^4P_{1/2}$ - $^2P^o_{1/2}$) doublet of Si II. The two experimental measurements are leftmost followed by the theoretical values in reverse chronological order left to right. The horizontal line lies at the BR as measured in this work as a guide for the eye.

4. Silicon Abundances in Very Metal-Poor Stars

All but two of the transitions studied here lie in the ultraviolet (UV) spectral domain below the atmospheric absorption cutoff, i.e. $\lambda < 3000$ Å. This limits high-resolution stellar spectroscopy to the Space Telescope Imaging Spectrograph (STIS; Kimble et al. 1998;

Woodgate et al. 1998), on board the Hubble Space Telescope (HST). Additionally, the UV spectrum is crowded with strong absorption lines of light and Fe-group elements, making reliable abundance analyses difficult to execute. The UV spectral region of cool stars features complex blends of transitions with various pedigrees, ranging from prominent well-known lines that have well-documented laboratory histories to many moderate and weak lines with poor or completely unknown atomic parameters.

The UV lines of neutral Si studied here are almost all very strong, having low excitation energies ($\chi$ < 6299 cm$^{-1}$ or < 0.8 eV) and relatively large transition probabilities (17 out of 20 lines in Table 3 have log($gf$) > –3). The problem here is not in identifying Si I lines; it is in finding stars with lines that are weak enough for abundance analysis. With this unusual constraint we concentrated on metal-poor ([Fe/H] < –2) halo stars that have been observed by HST/STIS. The list is small: 7 stars are considered in the metallicity study of Roederer et al. (2018); the bright main sequence star HD 84937 ([Fe/H] ≈ –2.2) has been featured in previous papers in this series (Den Hartog et al. 2021, and references therein); the famous warm low metallicity stars HD 19445 and HD 140283 (Chamberlain & Aller 1951) have been featured in several UV line identification contributions (e.g., Peterson et al. 2020 and references therein); the mildly metal-poor warm giant HD 222925 ([Fe/H] = -1.5) has been recently studied by Roederer et al. (2022) to produce a nearly complete abundance set for 63 elements. A few other such stars can be found but do not change the basic results which we will discuss here.

We employed HST/STIS spectra of seven of the stars included in the papers cited above in order to explore if Si abundances derived from UV spectra could be more trustworthy than the few optical-wavelength lines treated in the literature. We supplemented our HST/STIS spectra with blue spectra collected using the High Resolution Echelle Spectrometer (Vogt et al. 1994) at the Keck I telescope, and the Ultraviolet and Visual Echelle Spectrograph (Dekker et al. 2000) at the Very Large Telescope. We accessed these data through the Keck Observatory Archives and European Southern Observatory Archives, respectively, and Table 1 of Roederer et al. (2018) presents a description of these data.

We derived Si abundances using synthetic/observed spectrum matches. The synthetic spectra were computed with the plane-parallel LTE (local thermodynamic equilibrium) line analysis code MOOG (Sneden 1973)[12]. Atomic line lists for these syntheses were generated with the *linemake* facility (Placco et al. 2021)[13], which emphasizes laboratory transition data on Fe-group and neutron-capture neutral and singly-ionized species from the Wisconsin atomic physics group and on molecular species from the Old Dominion University group (e.g., Brooke et al. 2016, and references therein). We adopted the atmosphere parameters of Roederer et al. 2018, 2022) to produce model atmospheres interpolated from the ATLAS grid (Kurucz 2011, 2018)[14].

---

[12] Available at https://www.as.utexas.edu/~chris/moog.html
[13] https://github.com/vmplacco/linemake
[14] http://kurucz.harvard.edu/grids.html

For almost all stars the lower wavelength boundary of our HST/STIS spectra was $\lambda \approx$ 2300 Å, thus ruling out work on the five lowest-wavelength Si I transitions.

Our initial synthetic spectrum tests yielded results that further narrowed the range of stellar parameters that are useful for this abundance exercise. For stars that have [Fe/H] > –2.5 and effective temperatures $T_{eff}$ < 6000 K, many of the promising Si I lines simply are too strong and/or too blended with other strong neutral and ionized species features to yield reliable abundances. In particular, we discarded the giant star HD 222925 ($T_{eff}$/log($g$)/[M/H]/$v_t$ = 5636K/2.54/–1.5/2.20km s$^{-1}$; Roederer et al. 2022) and the subgiant HD 140283 (5600K/3.66/-2.6/1.15km s$^{-1}$; Roederer et al. 2018). We report here on five very metal-poor main sequence turnoff stars that have $T_{eff} \geq$ 6050 K.

In Table 5 we list the model parameters, individual line abundances, and final species abundances for both Si I and Si II transitions in the program stars. The mean abundances are based on 10-11 Si I lines and 2 Si II lines, all in the vacuum UV spectral domain, whereas in previous studies the Si abundances of these kinds of stars have come almost exclusively from the optical Si I transitions at 3905.5 and 4102.9 Å. We derive <[Si/Fe]$_I$> = 0.43 ($\sigma$ = 0.11). The inclusion of the ionized species in Si abundance studies is a rarity, and for our program stars the abundance agreement between neutral and ion is excellent. From Table 5 we find <[Si/Fe]$_{II}$ – [Si/Fe]$_I$> = +0.03 ($\sigma$ = 0.05). In Figure 5 we show small spectral regions around both Si II lines and around six representative Si I lines in the program star BD+03º 740. For this star and the other two lowest metallicity stars BD-13º 3442 and CD-33º 1173 the Si II lines are essentially on the weak-line linear part of the curve of growth. They are easy to detect, and to employ in abundance analyses. Many Si I lines are also reliable abundance indicators. However, the 2516, 2519, and 2881 Å transitions are clearly saturated and thus less sensitive to abundance changes. In cooler, higher metallicity stars such as HD 19445 and HD 84937 these and other lines become so strong that they are untrustworthy for abundance determinations. Some caution should be used in interpreting the Si abundances of those stars.

We also derived abundances for the Si I 3905 Å line and list them in Table 5. The 4102.9 Å Si I line was too weak and too blended with the strong H$\delta$ 4101.75 Å feature in our stars. However we did not include the 3905 Å line in the mean abundance calculations because this transition is known to yield temperature-dependent abundances in LTE calculations. Si in metal-poor giants from the $\lambda$3905 line is almost uniformly overabundant, <[Si/Fe]> ~ +0.4 ± 0.1 (e.g, Cayrel et al. 2004), but is much less abundant in main sequence stars near the turnoff region, <[Si/Fe]> ~ +0.1 ± 0.1 (e.g, Cohen et al. 2004). The sample of horizontal-branch stars investigated by Preston et al. (2006) covers a large temperature range and shows this effect clearly in their Figure 8. A summary of the observational issues in LTE abundances was discussed by Sneden & Lawler (2008). From Table 5 we compute <[Si/Fe]> = +0.28 ($\sigma$ = 0.11) from the 3905 Å line, clearly lower than the mean from the UV Si I lines discussed above. Amarsi & Asplund (2017) computed NLTE corrections for optical-wavelength Si I transitions in the solar photosphere, and have published on-line tables of NLTE corrections for many ($T_{eff}$/log($g$)/[Fe/H]/$v_t$) combinations.[15] Their suggested correction for the 3905 Å line in stars with parameters (6000 K/4.0/-3.0/1-2 km s$^{-1}$) is Δ[Si/Fe] ≅ +0.1 dex. Applying this adjustment to the abundances from this line for our stars would bring the 3905 Å line into better agreement with our abundances derived from the UV Si I transitions. Abundances from the UV lines should be preferred.

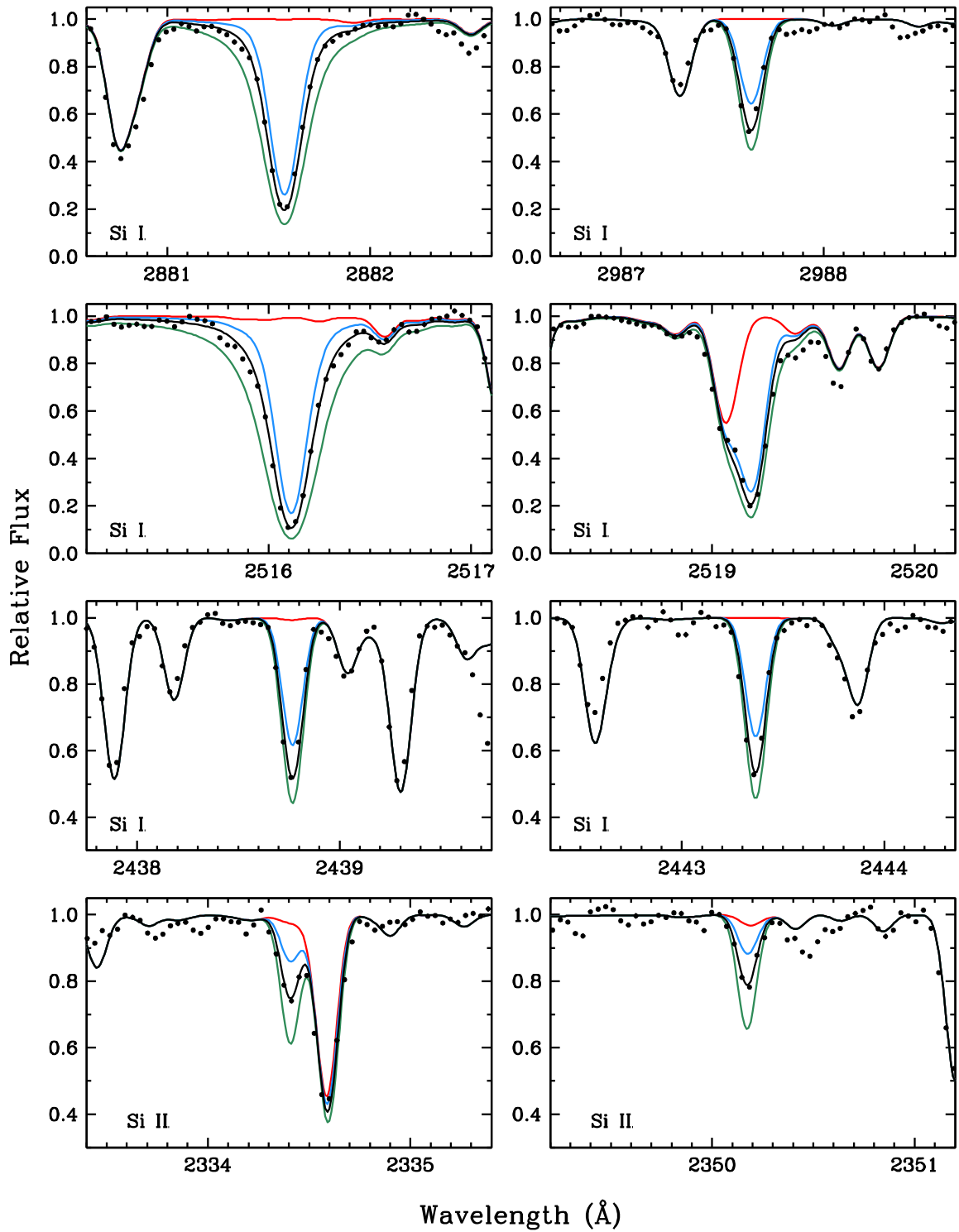

Figure 5: Observed and synthesized spectra for both Si II lines (the 2 bottom panels) and for representative lines of Si I (the 6 upper panels) in the star BD+03º 740. In each panel, the filled circles are the observations. The red line is a synthesis without any contribution from Si. The best fit to a line is given by the black line, and the blue and green lines show the synthetic spectra for Si abundances 0.4 dex lesser and greater than the best match.

5. Discussion

In Figure 6 we illustrate the Galactic Chemical Evolution (GCE) trends of [Si/Fe] as a function of metallicity ([Fe/H]). Silicon is synthesized in explosive oxygen burning, and thus is formed in core-collapse supernovae early in the history of the Galaxy and then ejected into the gas that eventually forms the halo stars. (Curtis et al. 2019) We show a compilation of abundance data, [Si/Fe], from an earlier survey of low-metallicity Galactic stars (Roederer et al. 2014; shown as open squares). The values of [Si/Fe] exhibit significant scatter over the observed metallicity range. This could be the result of comparing different types of stars (i.e., dwarfs with giants) or due to the choice of the atomic lines used for the abundance determinations and/or the source of the log(*gf*)s employed. Employing our new experimental silicon data (discussed above, see Tables 3 and 4) leads to a more consistent pattern with less scatter. For the five stars in this study (shown as filled red circles in Figure 6) the average value of [Si/Fe] = 0.44, significantly higher than the solar value of 0. This value can serve as a constraint on GCE models and, in particular, on supernovae nucleosynthesis model predictions for early Galactic times.

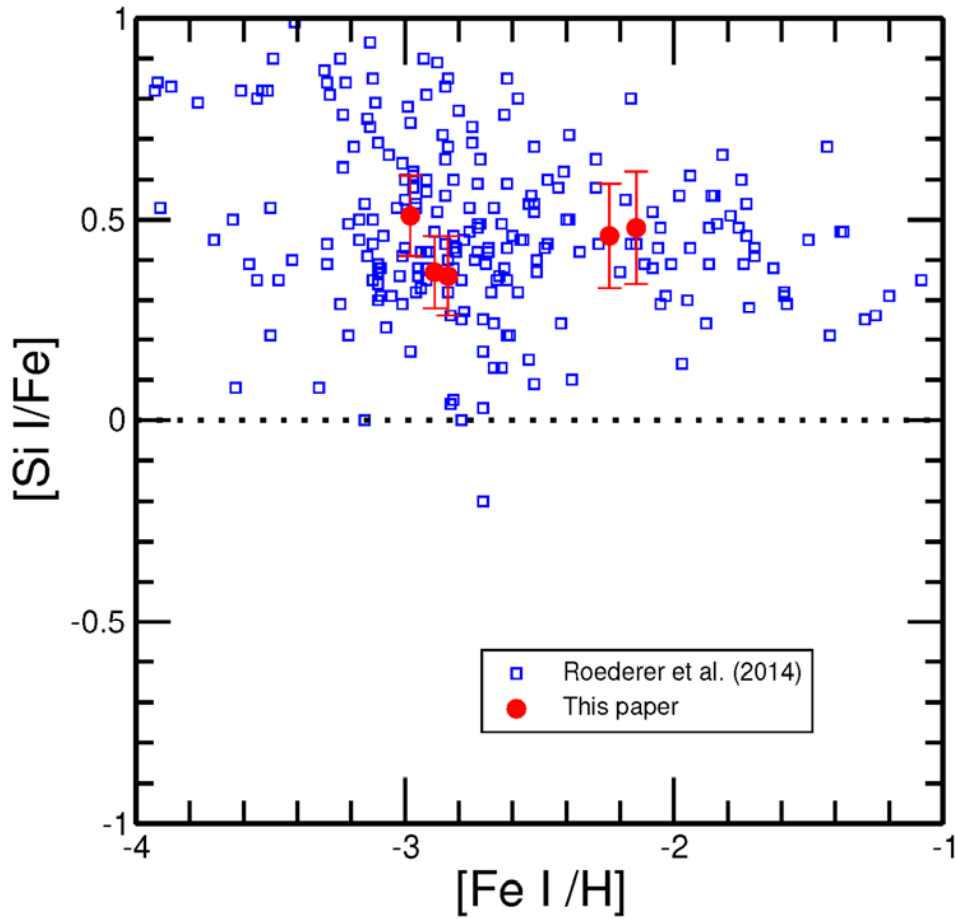

Figure 6. The [Si/Fe] abundance ratios as a function of metallicity ([Fe/H]) for metal-poor stars from Roederer et al. (2014) (blue open squares) and this paper (red filled circles).

---

[15] Anish Amarsi - Theoretical Astrophysics, Department of Physics and Astronomy, Uppsala University - Astronomy and Space Physics Theoretical astrophysics Department of Physics and Astronomy Uppsala University Box 516, 75120 Uppsala Sweden; Email: anish.amarsi@physics.uu.se ; www.astro.uu.se

It would be expected that the [Si/Fe] values illustrated in Figure 6 would begin to exhibit a downward pattern at metallicities closer to [Fe/H] = -1 with the onset of Type Ia supernovae (the main producer of iron) throughout the Galaxy. The abundance data from Roederer et al. (2014) does hint at such a downward trend, but clearly more studies employing the new precise atomic data in somewhat more metal-rich stars will be needed to confirm such a trend.

## 6. Conclusions

We have made new BF measurements for 20 UV and blue lines of Si I as well as the $^4P_{1/2}$ intercombination lines of Si II. Comparisons are made to earlier experiment as well as theory. These BF have been combined with radiative lifetimes measured previously to determine *A*-values and log(*gf*)s for these transitions. The current study represents a significant improvement in measurement of the very weak spin-forbidden lines of both Si I and Si II. These new data have been applied to abundance determinations in five metal-poor main sequence turnoff stars. We find that many of the Si I UV transitions can be used as reliable abundance indicators in very metal-poor stars and we obtain excellent agreement between abundances determined using Si I transitions and the Si II intercombination lines.

## ACKNOWLEDGEMENTS

This work is supported by NSF grant AST-1814512 and AST-2206050 (E.D.H. and J.E.L). I.U.R. acknowledges support from NSF grants AST 2205847 and PHY 14-30152 (Physics Frontier Center/JINA-CEE), and NASA grants GO-14232, GO-15657 and AR-16630 from the Space Telescope Science Institute, which is operated by the Association of Universities for Research in Astronomy, Incorporated, under NASA contract NAS5-26555. We are grateful to Hampus Nilsson for sharing the Si I, II data from the Pehlivan Rhodin (2018) thesis prior to its publication, and to Karen Lind for helpful discussions.

Facilities: HST (STIS), Keck I (HIRES), VLT (UVES).

Software: LINEMAKE (Placco et al. 2021), MOOG (Sneden 1973).

Table 1. Echelle spectra of commercial HCLs used in the study of Si II and Si I BFs.[a]

| Index[b] | Date | Serial Number | Buffer Gas | Lamp Current (mA) | Spectral Coverage (Å) | Coadds | Total Exposure (min) |
|---|---|---|---|---|---|---|---|
| 11 | 2021 Jul 31 | 2 | Neon | 12 | 2090 - 2955 | 120 | 360 |
| 12 | 2021 Aug 03 | 2 | Neon | 12 | 2090 - 2955 | 5 | 150 |
| 13 | 2021 Aug 12 | 1 | Neon | 12 | 2090 - 2955 | 6 | 180 |
| 14 | 2021 Aug 27 | 1 | Neon | 20 | 2090 - 2955 | 15 | 150 |
| 15 | 2021 Aug 29 | 1 | Neon | 22 | 2040 - 2790 | 18 | 180 |
| 16 | 2021 Sept 04 | 1 | Neon | 15 | 2040 - 2790 | 3 | 180 |
| 17 | 2022 Apr 02 | 1 | Neon | 25 | 2040 - 2790 | 4 | 120 |
| 18 | 2022 Apr 05 | 1 | Neon | 25 | 2040 - 2790 | 16 | 160 |
| 31 | 2021 Dec 17 | 1 | Neon | 12 | 2150 - 3245 | 80 | 120 |
| 32 | 2021 Dec 17 | 3 | Neon | 12 | 2350 - 4300 | 144 | 120 |
| 33 | 2022 Jan 06 | 3 | Neon | 18 | 2150 - 3245 | 90 | 60 |
| 34 | 2022 Jan 06 | 1 | Neon | 18 | 2350 - 4300 | 120 | 60 |
| 35 | 2022 Jan 06 | 5 | Neon | 6 | 2150 - 3245 | 15 | 75 |
| 36 | 2022 Jan 06 | 7 | Neon | 6 | 2350 - 4300 | 37 | 74 |
| 37 | 2022 Jan 08 | 1 | Neon | 12 | 2350 - 4300 | 720 | 120 |
| 38 | 2022 Jan 08 | 3 | Neon | 18 | 2350 - 4300 | 90 | 90 |
| 39 | 2022 Jan 08 | 5 | Neon | 6 | 2350 - 4300 | 60 | 60 |
| 40 | 2022 Jan 13 | 1 | Neon | 24 | 2350 - 4300 | 120 | 60 |
| 41 | 2022 Jan 13 | 3 | Neon | 29 | 2350 - 4300 | 120 | 60 |
| 42 | 2022 Mar 29 | 1 | Neon | 24 | 2350 - 4300 | 240 | 60 |
| 43 | 2022 Mar 29 | 3 | Neon | 28 | 2350 - 4300 | 240 | 60 |
| 44 | 2022 Apr 05 | 3 | Neon | 25 | 2350 - 4300 | 180 | 60 |
| 45 | 2022 Apr 07 | 1 | Neon | 12 | 2280 – 4200 | 60 | 60 |
| 46 | 2022 Apr 07 | 3 | Neon | 18 | 2280 - 4200 | 90 | 60 |
| 47 | 2022 Apr 07 | 5 | Neon | 25 | 2150 – 3245 | 103 | 60 |
| 48 | 2022 May 14 | 1 | Argon | 18 | 2050 – 2800 | 100 | 100 |
| 49 | 2022 May 19 | 1 | Argon | 20 | 2050 – 2800 | 60 | 60 |
| 50 | 2022 May 19 | 3 | Argon | 20 | 2150 – 3245 | 60 | 60 |
| 51 | 2022 May 19 | 5 | Neon | 28 | 2150 – 3245 | 144 | 60 |
| 52 | 2022 May 21 | 1 | Neon | 30 | 2150 - 3245 | 240 | 60 |
| 53 | 2022 May 21 | 3 | Argon | 19 | 2350 - 4300 | 240 | 60 |
| 54 | 2022 May 22 | 1 | Neon | 21 | 2350 -4300 | 120 | 120 |
| 55 | 2022 May 30 | 1 | Neon | 3 | 2350 - 4300 | 24 | 120 |
| 56 | 2022 May 30 | 3 | Neon | 3 | 2280 - 4200 | 24 | 120 |

Note:
[a] All echelle spectra were taken from commercially manufactured Si-Ne or Si-Ar HCLs, and have a spectral resolving power of ~250,000 although the effective resolving power is somewhat lower due to line broadening. Each of the spectra were calibrated with a $D_2$ lamp spectrum, which was recorded immediately following the completion of each HCL spectrum. Each spectrum listed is a single CCD frame, and does not cover an entire echelle grating order, but is sufficient coverage to determine branching fractions of all transitions from one or more upper levels studied.
[b] The first eight spectra list (indices 11 – 18) were used to study the BF of the Si II intercombination lines. The remaining spectra (indices 31 – 56) were used in the Si I BF study.

Table 2.
Branching Fractions of Si I

| Upper level[a] | | Lower level[a] | | $\lambda_{air}$ | $\sigma_{vac}$ | This Expt. | | Other Expt.[b] | | LS[c] |
|---|---|---|---|---|---|---|---|---|---|---|
| Configuration and Term | $E_k$ (cm$^{-1}$) | Term[d] | $E_i$ (cm$^{-1}$) | (Å) | (cm$^{-1}$) | BF | (±%) | BF | (±%) | BF |
| $3s^23p4s$ $^3P°_1$ | 39760.285 | $^3P_0$ | 0.000 | 2514.316 | 39760.20 | 0.337 | (1) | 0.333 | (0.9) | 0.333 |
| | | $^3P_1$ | 77.115 | 2519.202 | 39683.17 | 0.244 | (1) | 0.247 | (1.6) | 0.248 |
| | | $^3P_2$ | 223.157 | 2528.508 | 39537.11 | 0.409 | (1) | 0.407 | (1.0) | 0.409 |
| | | $^1D_2$ | 6298.850 | 2987.643 | 33461.42 | 0.0103 | (6) | 0.012 | (8) | |
| | | $^1S_0$ | 15394.370 | 4102.936 | 24365.91 | 0.00056 | (17) | <0.0020 | (30) | |
| $3s^23p4s$ $^3P°_2$ | 39955.053 | $^3P_1$ | 77.115 | 2506.897 | 39877.90 | 0.243 | (1) | 0.246 | (1.2) | 0.252 |
| | | $^3P_2$ | 223.157 | 2516.112 | 39731.88 | 0.757 | (1) | 0.754 | (0.4) | 0.748 |
| | | $^1D_2$ | 6298.850 | 2970.353 | 33656.18 | 0.00020 | (10) | 0.00027 | (13) | |
| $3s^23p4s$ $^1P°_1$ | 40991.884 | $^3P_0$ | 0.000 | 2438.768 | 40991.80 | 0.0030 | (7) | 0.0034 | (5.9) | |
| | | $^3P_1$ | 77.115 | 2443.365 | 40914.80 | 0.0024 | (7) | 0.0027 | (7.4) | |
| | | $^3P_2$ | 223.157 | 2452.118 | 40768.70 | 0.0022 | (7) | 0.0025 | (8.7) | |
| | | $^1D_2$ | 6298.850 | 2881.578 | 34693.02 | 0.940 | (0.5) | 0.934 | (0.2) | |
| | | $^1S_0$ | 15394.370 | 3905.523 | 25597.51 | 0.052 | (9) | 0.057 | (2.2) | |
| $3s3p^3$ $^3D°_1$ | 45276.188 | $^3P_0$ | 0.000 | 2207.978 | 45276.10 | 0.566 | (1) | 0.577 | (1.4) | 0.557 |
| | | $^3P_1$ | 77.115 | 2211.745 | 45199.20 | 0.409 | (1) | 0.398 | (2.3) | 0.415 |
| | | $^3P_2$ | 223.157 | 2218.916 | 45053.10 | 0.025 | (4) | 0.023 | (13) | 0.027 |
| | | $^1D_2$ | 6298.850 | 2564.825 | 38977.34 | 0.00044 | (26) | <0.000 | (15) | |
| $3s3p^3$ $^3D°_2$ | 45293.629 | $^3P_1$ | 77.115 | 2210.892 | 45216.60 | 0.763 | (0.5) | 0.760 | (0.4) | 0.751 |
| | | $^3P_2$ | 223.157 | 2218.057 | 45070.40 | 0.236 | (1) | 0.240 | (1.3) | 0.248 |
| | | $^1D_2$ | 6298.850 | 2563.679 | 38994.78 | 0.00053 | (19) | <0.000 | (5) | |

Notes:
[a] Upper and lower levels are taken from NIST ASD and are ordered by term.
[b] Sm87: Smith et al. 1987, ApJ 322, 573. The BFs of weak lines at 2438.768 Å and 2443.365 Å have only one significant digit listed in Sm87 Table 1. We have calculated to two significant digits from their log(gf)s.
[c] LS BFs within the triplet multiplets calculated from the relative line strengths in Appendix I of Cowan (1981) and with frequency-cubed scaling. They are renormalized to the total multiplet strength from the current measurements.
[d] The configuration of all lower levels is $3s^23p^2$

Table 3.
A-values and log(gf)s for 20 transitions of Si I

| $\lambda_{air}$ | $E_{upper}$ | $J_{upper}$ | $E_{lower}$ | $J_{lower}$ | This Expt. | | | Sav16[a] | PR18 | |
|---|---|---|---|---|---|---|---|---|---|---|
| (Å) | (cm$^{-1}$) | | (cm$^{-1}$) | | $A_{ki}$ (s$^{-1}$) | (±%) | log(gf) | log(gf) | log(gf) | (±%) |
| 2207.978 | 45276.188 | 1 | 0.000 | 0 | 2.57E+07 | (5) | -1.248 | -1.229 | -1.318 | (6.5) |
| 2210.892 | 45293.629 | 2 | 77.115 | 1 | 3.47E+07 | (5) | -0.895 | -0.876 | -0.965 | (7.5) |
| 2211.745 | 45276.188 | 1 | 77.115 | 1 | 1.86E+07 | (5) | -1.388 | -1.372 | -1.459 | (5.8) |
| 2218.057 | 45293.629 | 2 | 223.157 | 2 | 1.07E+07 | (5) | -1.402 | -1.392 | -1.477 | (6) |
| 2218.916 | 45276.188 | 1 | 223.157 | 2 | 1.13E+06 | (6) | -2.603 | -2.586 | -2.670 | (4.4) |
| 2438.768 | 40991.884 | 1 | 0.000 | 0 | 7.06E+05 | (9) | -2.723 | -2.684 | -2.705 | (5.6) |
| 2443.365 | 40991.884 | 1 | 77.115 | 1 | 5.52E+05 | (9) | -2.828 | -2.788 | -2.805 | (0.7) |
| 2452.118 | 40991.884 | 1 | 223.157 | 2 | 5.01E+05 | (9) | -2.868 | -2.829 | -2.850 | (3.4) |
| 2506.897 | 39955.053 | 2 | 77.115 | 1 | 5.39E+07 | (5) | -0.595 | -0.566 | -0.578 | (0.9) |
| 2514.316 | 39760.285 | 1 | 0.000 | 0 | 7.48E+07 | (5) | -0.672 | -0.667 | -0.679 | (0.8) |
| 2516.112 | 39955.053 | 2 | 223.157 | 2 | 1.68E+08 | (5) | -0.098 | -0.088 | -0.101 | (0.9) |
| 2519.202 | 39760.285 | 1 | 77.115 | 1 | 5.42E+07 | (5) | -0.810 | -0.793 | -0.805 | (0.9) |
| 2528.508 | 39760.285 | 1 | 223.157 | 2 | 9.08E+07 | (5) | -0.583 | -0.567 | -0.579 | (0.8) |
| 2563.679 | 45293.629 | 2 | 6298.850 | 2 | 2.43E+04 | (20) | -3.922 | -3.953 | -4.078 | (25.2) |
| 2564.825 | 45276.188 | 1 | 6298.850 | 2 | 2.00E+04 | (26) | -4.228 | -4.298 | -4.389 | (28.7) |
| 2881.578 | 40991.884 | 1 | 6298.850 | 2 | 2.19E+08 | (5) | -0.088 | -0.044 | -0.061 | (1.4) |
| 2970.353 | 39955.053 | 2 | 6298.850 | 2 | 4.44E+04 | (11) | -3.531 | -3.577 | -3.613 | (6.4) |
| 2987.643 | 39760.285 | 1 | 6298.850 | 2 | 2.30E+06 | (8) | -2.035 | -2.082 | -2.113 | (3) |
| 3905.523 | 40991.884 | 1 | 15394.370 | 0 | 1.22E+07 | (10) | -1.077 | -0.999 | -1.018 | (3.5) |
| 4102.936 | 39760.285 | 1 | 15394.370 | 0 | 1.24E+05 | (18) | -3.026 | -3.126 | -3.154 | (3) |

Notes
[a] log(gf)s calculated from A-values presented in Sav16 using equation 2.

Table 4.
Branching Fractions, *A*-values and log(*gf*)s for the $^4P_{1/2}$ - $^2P_{1/2,3/2}$ doublet of Si II from experiment and recent theory.

| | $3s3p^2\ ^4P_{1/2} - 3s^23p\ ^2P°_{1/2}$; $\lambda_{air}$ = 2334.407 Å | | | $3s3p^2\ ^4P_{1/2} - 3s^23p\ ^2P°_{3/2}$; $\lambda_{air}$ = 2350.172 Å | | |
|---|---|---|---|---|---|---|
| | BF | *A* (s$^{-1}$) | log(*gf*) | BF | *A* (s$^{-1}$) | log(*gf*) |
| This Expt:[a] | 0.519 ± 1% | 4990 ± 16% | -5.088 | 0.481 ± 1% | 4630 ± 16% | -5.116 |
| Other Expt: CSB93 | 0.541 ± 10% | 5200 ± 19% | -5.070 | 0.459 ± 10% | 4410 ± 21% | -5.136 |
| Theory: PR18[b] | 0.520 | 5280 ± 18.9% | -5.064 | 0.480 | 4882 ± 11.7% | -5.092 |
| Theory: Wu20[b] | 0.514 | 5230 | -5.068 | 0.486 | 4940 | -5.087 |

Notes:
[a] our BFs are combined with the lifetime of CSB93 (104 ± 16 μs) to determine our *A*-value and log(*gf*)
[b] PR18 and Wu20 do not report BFs. We calculate BFs from their *A*-values to show the excellent agreement with the BFs measured in this study.

Table 5.
Line-by-line abundances from Si I and Si II lines for the five metal-poor stars investigated.

| | | Stellar Parameters | | | | | |
|---|---|---|---|---|---|---|---|
| | | star | BD+03º 740 | BD-13º 3442 | CD-33º 1173 | HD 19445 | HD 84937 |
| | | $T_{eff}$ | 6351 | 6405 | 6625 | 6055 | 6300 |
| | | $\log(g)$ | 3.97 | 4.04 | 4.29 | 4.49 | 4.00 |
| | | $v_t$ | 1.7 | 1.6 | 1.6 | 1.2 | 1.5 |
| | | [Fe I/H] | -2.89 | -2.84 | -2.98 | -2.14 | -2.24 |
| | | [Fe II/H] | -2.78 | -2.73 | -2.90 | -2.17 | -2.26 |
| | | source | Cowan20 | Cowan20 | Cowan20 | Roederer18 | Sneden16 |
| | | Line-by-Line Abundances – Si I | | | | | |
| $\lambda$ (Å) | $\chi$ (eV) | $\log(gf)$ | [Si I/Fe] | [Si I/Fe] | [Si I/Fe] | [Si I/Fe] | [Si I/Fe] |
| 2438.768 | 0.000 | -2.723 | 0.34 | 0.29 | 0.48 | 0.64 | 0.57 |
| 2443.365 | 0.010 | -2.828 | 0.37 | 0.34 | 0.48 | 0.59 | 0.47 |
| 2452.118 | 0.028 | -2.868 | 0.41 | 0.34 | 0.43 | … | 0.42 |
| 2506.897 | 0.010 | -0.595 | 0.44 | 0.46 | 0.58 | 0.64 | 0.52 |
| 2514.316 | 0.000 | -0.672 | 0.49 | 0.46 | 0.58 | 0.59 | 0.52 |
| 2516.112 | 0.028 | -0.098 | 0.49 | 0.49 | 0.63 | 0.49 | 0.52 |
| 2519.202 | 0.010 | -0.810 | 0.34 | … | 0.58 | … | 0.57 |
| 2528.508 | 0.028 | -0.583 | 0.34 | 0.49 | 0.58 | … | 0.57 |
| 2564.825 | 0.780 | -3.922 | … | … | … | 0.39 | … |
| 2881.578 | 0.780 | -0.088 | 0.24 | 0.29 | 0.33 | 0.39 | 0.32 |
| 2970.353 | 0.780 | -3.531 | 0.41 | 0.24 | … | 0.29 | 0.37 |
| 2987.643 | 0.780 | -2.035 | 0.19 | 0.24 | 0.38 | 0.29 | 0.17 |
| *3905.523[a]* | *1.907* | *-3.026* | *0.14* | *0.24* | *0.28* | *0.44* | *0.32* |
| | | mean[a] | 0.37 | 0.36 | 0.51 | 0.48 | 0.46 |
| | | $\sigma$ | 0.09 | 0.10 | 0.10 | 0.14 | 0.13 |
| | | Line-by-Line Abundances – Si II | | | | | |
| $\lambda$ (Å) | $\chi$ (eV) | $\log(gf)$ | [Si II/Fe] | [Si II/Fe] | [Si II/Fe] | [Si II/Fe] | [Si II/Fe] |
| 2334.407 | 0.036 | -5.088 | 0.43 | 0.48 | 0.50 | 0.57 | 0.52 |
| 2350.172 | 0.036 | -5.116 | 0.36 | 0.43 | 0.65 | 0.32 | 0.37 |
| | | mean | 0.40 | 0.46 | 0.58 | 0.45 | 0.45 |
| | | $\sigma$ | 0.05 | 0.04 | 0.11 | 0.18 | 0.11 |

Note
[a]The mean and standard deviation of [Si I/Fe] are calculated without the λ3905 line data as this this transition is known to yield temperature-dependent abundances in LTE calculations. See text for further discussion.